\documentclass[10pt, conference, letterpaper]{IEEEtran}

\usepackage{amsmath}
\usepackage{graphicx}
\usepackage{hyperref}
\usepackage{cite}
\usepackage{url}
\usepackage{multirow}
\bstctlcite{IEEEexample:BSTcontrol}

\usepackage{amsfonts}

\usepackage{xcolor}
\newcommand{\new}[1]{\textcolor{black}{#1}}

\title{MambaLiteSR: Image Super-Resolution with Low-Rank Mamba using Knowledge Distillation}

\author{\IEEEauthorblockN{Romina Aalishah, Mozhgan Navardi, and Tinoosh Mohsenin}
\IEEEauthorblockA{Department of Electrical and Computer Engineering\\
Johns Hopkins University\\
\{raalish1, mnavard1, tinoosh\}@jhu.edu}
}

\begin{document}

\maketitle

\begin{abstract}
Generative Artificial Intelligence~(AI) has gained significant attention in recent years, revolutionizing various applications across industries. Among these, advanced vision models for image super-resolution are in high demand, particularly for deployment on edge devices where real-time processing is crucial. However, deploying such models on edge devices is challenging due to limited computing power and memory. In this paper, we present MambaLiteSR, a novel lightweight image Super-Resolution~(SR) model that utilizes the architecture of Vision Mamba. It integrates State Space Blocks and a reconstruction module for efficient feature extraction. To optimize efficiency without affecting performance, MambaLiteSR employs knowledge distillation, transferring essential information from a larger Mamba-based teacher model to a smaller student model through hyperparameter tuning. Through a mathematical analysis of model parameters and their impact on the Peak Signal-to-Noise Ratio~(PSNR), we identify key factors and adjust them accordingly. Our comprehensive evaluation shows that MambaLiteSR outperforms state of the art edge SR methods by reducing power consumption while maintaining competitive PSNR and SSIM scores across benchmark datasets such as Set5, Set14, and BSD100. It also reduces the power usage during training by adopting low-rank approximation. Moreover, MambaLiteSR reduces the total number of parameters without degrading performance, enabling the efficient deployment of generative AI models on resource-constrained devices. Deployment on the embedded NVIDIA Jetson Orin Nano confirms the superior balance of MambaLiteSR size, latency, and resource efficiency. The experimental results show that MambaLiteSR achieves performance comparable to both the baseline and other edge models while using 15\% fewer parameters than the baseline. It also improves the power consumption by up to \new{58\%} compared to state-of-the-art SR edge models, all while maintaining low energy consumption during training.
\end{abstract}

\begin{IEEEkeywords}
Image Super-Resolution, Mamba, Knowledge Distillation, Low-Rank Approximation, Edge Computing
\end{IEEEkeywords}

\section{Introduction}
Generative Artificial Intelligence~(AI) models have gained attention in recent years for their ability to generate outputs such as images and text based on input data~\cite{genai, transformer, llmaug}. These models learn the patterns and structures of training data and use them to produce results. Image Super-Resolution~(SR) is one such task in Generative AI and computer vision. The goal is to generate a High-Resolution~(HR) image from a Low-Resolution~(LR) one~\cite{yu2023review}. An efficient SR model achieves high Peak Signal-to-Noise Ratio~(PSNR) and Structural Similarity~(SSIM); PSNR measures the reconstruction quality by comparing the pixel-wise difference between the HR output and the ground truth, while SSIM evaluates the perceptual similarity by considering structural information, luminance, and contrast. However, deploying these models on edge devices might be challenging because of their complexity and high intensiveness~\cite{fatrabbit, kallakuri2024resource, navardi2024metatinyml, rashid2024tinym, bao2025decentralised, reprohrl, mlae2}. Since edge devices are constrained in several aspects such as computing resources and power, the importance of model size, number of parameters, and FLOPs comes into matter~\cite{navarrete2022edge, wang2024eshp, mazumder2024reg, meta, yolo}. 

\new{Deep learning models play a crucial role in applications such as image processing~\cite{water}. Two commonly used architectures in deep learning are Fully Connected Networks (FCNs) and Convolutional Neural Networks (CNNs).} Early SR approaches relied on CNNs~\cite{srcnn, vdsr, fsrcnn}. While small, these CNN-based methods demonstrated relatively low PSNRs. Therefore, efforts have been made to improve the performance~\cite{rcan, edsr, denoiser} by removing unnecessary modules, bypassing low-frequency information, and integrating denoising as a prior. Recently, research has focused on developing smaller and faster models with novel architectures, attention mechanisms, and hybrid models~\cite{eshr, esr, mafdn}.

With the introduction of Mamba~\cite{mamba}, a revolution happened in optimizing feature extraction. This architecture was quickly adopted across various vision tasks, including image classification, object detection, and semantic segmentation~\cite{visionmamba, mambavision}. Having similar results to previous works but with better processing speed, led to its adoption in SR frameworks, where it involved combining the lightweight structure of Mamba with attention mechanisms and transformer-based modules~\cite{lei2024dvmsr, himamba, srmamba}. However, even though Mamba improved the latency, model deployment on edge devices is still challenging because of its size and power usage.

To address these challenges, we propose MambaLiteSR, which builds on previous studies by enhancing the existing Mamba-based architecture with significant improvements. In this paper, we focus on minimizing the model size and power usage during inference and training while maintaining comparable performance by optimizing the loss function, \new{utilizing low-rank Mamba, knowledge distillation, and carefully fine-tuning hyperparameters}. After finalizing the architecture and loss function, we employed low-rank approximation, which is widely used in machine learning studies such as LoRA~\cite{lora}, to decrease the power usage during training. Then, knowledge distillation was applied to reduce the model size. To the best of our knowledge, no existing work has explored the simultaneous use of Mamba, knowledge distillation, and low-rank approximation for optimizing SR generative AI models on edge devices. In this context, we enhanced the Vision~Mamba-based model to improve performance, energy usage while training, and compactness, ensuring it fits seamlessly on edge devices. To evaluate the proposed MambaLiteSR, experiments were conducted on key knowledge distillation parameters to achieve optimal results. The experimental results demonstrate that MambaLiteSR achieves comparable performance to baseline and edge models while being 15\% smaller than the baseline. For edge-device evaluation, the model was tested on the embedded NVIDIA Jetson Orin Nano~\cite{orin}. The results show that our model achieves comparable PSNR and SSIM with the state-of-the-art edge models while consuming less dynamic power.
Our contributions for the proposed MambaLiteSR are summarized as follows:

\begin{itemize}
    \item Improvement of the image super-resolution model to achieve comparable performance by optimizing the loss function, utilizing knowledge distillation, and carefully fine-tuning its hyperparameter.
    \item Assessment of the matrix ranks used in Mamba architecture on model size, performance, and training power usage, as well as the distillation parameter $\alpha$ on model performance.
    \item Real-world experiment by deployment of the proposed MambaLiteSR on embedded NVIDIA Jetson Orin Nano, with comparisons made against similar models in terms of power usage and performance.
\end{itemize}

\section{Related Work}
Image super-resolution has been a major task in computer vision, with early approaches relying on interpolation techniques such as bicubic scaling~\cite{jahnavi2024study, navarrete2022edge, chen2015train}. While these methods are computationally efficient, they lack adaptability to image features and often produce artifacts like blurriness and jagged edges in images. Advances like SRCNN~\cite{srcnn} and FSRCNN~\cite{fsrcnn} introduced learnable features through CNNs, significantly improving SR performance. However, these models are computationally intensive and might be challenging for real-time deployment on resource-constrained devices.

To address efficiency concerns, lightweight SR models such as ESPCN~\cite{shi2016real}, DVMSR~\cite{lei2024dvmsr}, and SRMamba-T~\cite{srmamba} have been proposed. ESPCN utilizes sub-pixel convolution layers to reduce computational overhead, but it struggles to maintain high reconstruction quality at larger scales. DVMSR, which outperformed RLFN~\cite{rlfn}, the winner of the NTIRE 2023 Efficient Super-Resolution Challenge~\cite{ntire}, utilizes Vision Mamba~\cite{zhu2024visionmamba} modules and state space blocks to balance efficiency and accuracy. SRMamba-T combines Mamba and Transformer architectures to balance computational efficiency with high performance.

Frameworks like ESHP~\cite{wang2024eshp} extend these advancements by leveraging heterogeneous hardware to optimize SR tasks. ESHP dynamically allocates CPU, GPU, and NPU resources using deep reinforcement learning. However, its dependence on specialized hardware ecosystems introduces complexity and limits deployment flexibility. Edge-SR~\cite{navarrete2022edge} proposes lightweight one-layer architectures for real-time applications. Although practical for constrained devices, its performance is often inferior to more advanced multi-layer networks. Similarly, thermal imaging SR pipelines~\cite{mathur2021real} and facial verification systems~\cite{perez2023efficient} focus on specific use cases but are not flexible enough for wider SR applications.

The proposed MambaLiteSR further advances these foundations by integrating low-rank Mamba architecture and a knowledge distillation framework, offering a unified solution to the limitations of prior methods. Unlike simpler interpolation‐based techniques, MambaLiteSR dynamically adapts to complex image features for improved SR quality. Meanwhile, its design addresses the memory and computational constraints typically faced by CNN‐based or transformer‐heavy approaches, reducing reliance on specialized hardware. By managing parameter usage and supporting real‐time deployment, MambaLiteSR provides an effective, flexible solution for lightweight SR on edge devices.

\begin{figure}[t]
	\centering
	\includegraphics[width=.45\textwidth]
         {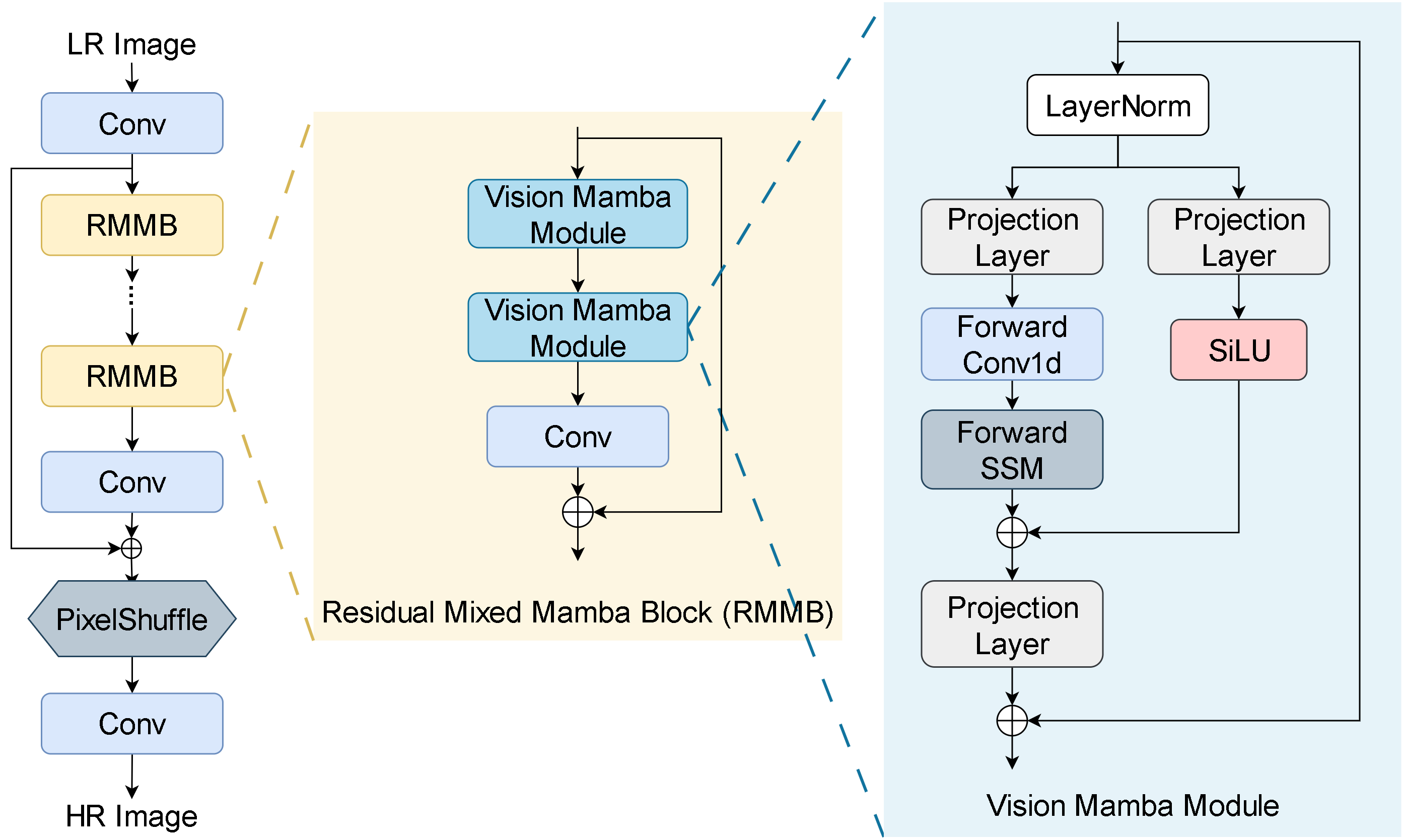}
	\caption{Architecture of DVMSR \cite{lei2024dvmsr} and Vision Mamba \cite{visionmamba}: Input image is preprocessed and fed into the DVMSR model, which consists of Vision Mamba modules, convolution layers, and a decoder.}
	\label{model}
\end{figure}

\section{Background}

Image super-resolution aims to reconstruct a high-resolution image from a low-resolution input while keeping details. Traditional methods cannot recover high-frequency details, leading to blurry results~\cite{freq, problem}. Deep learning-based SR models, particularly those utilizing CNN and Transformer-based architectures~\cite{srcnn, fsrcnn, srmamba}, have significantly improved performance. However, these models often require large computational resources, making deployment on edge devices challenging. In this section, various parts contributing to the reduction in model size and power usage are explored. In addition, strategies for improving the performance of small models are discussed.

\textbf{Mamba and State Space Models} The Mamba~\cite{mamba} architecture introduces Selective State Space Models, a lightweight alternative to Transformers, that reduces computational complexity while maintaining feature extraction capabilities. The state space representation in Mamba is formulated as follows:

\begin{equation}
    \mathbf{y}(t) = \mathbf{C} \mathbf{x}(t) 
\end{equation}
\begin{equation}
    \frac{d}{dt} \mathbf{x}(t) = \mathbf{A} \mathbf{x}(t) + \mathbf{B} \mathbf{u}(t)
\end{equation}

where $\mathbf{x}(t)$ represents the hidden state, $\mathbf{u}(t)$ is the input signal and $\mathbf{A}$, $\mathbf{B}$, and $\mathbf{C}$ are learnable matrices. This structure enables Mamba to capture long-range dependencies efficiently while requiring fewer parameters than traditional self-attention mechanisms. As a result, several efforts have been made to apply this method across various tasks. Vision Mamba~\cite{visionmamba} is such an effort that utilizes the efficient architecture of Mamba for computer vision tasks like image classification. Later, DVMSR~\cite{lei2024dvmsr} was proposed, utilizing Mamba Vision in the image super-resolution task. Figure~\ref{model} demonstrates the architecture of DVMSR and Vision Mamba, showing the contribution of Vision Mamba to DVMSR.


\textbf{Embedding Dimension} The embedding dimension significantly influences the number of parameters in the model. For an SR model, the number of parameters in a single layer can be estimated as:

\begin{equation}
    P = d_{in} \times d_{out} + d_{out}
\end{equation}

where $d_{in}$ and $d_{out}$ are the input and output embedding dimensions, respectively. Reducing the embedding dimension from $d_{base}$ to $d_{small}$ leads to a reduction in parameter count by a factor of:

\begin{equation}
    \frac{d_{small}^2}{d_{base}^2}
\label{eq:embed_dim}
\end{equation}




\textbf{Low-Rank Approximation} Low-rank approximation is used to further compress the model by reducing the rank of weight matrices while preserving essential information. Given a weight matrix $\mathbf{W} \in \mathbb{R}^{m \times n}$, a low-rank factorization approximates $\mathbf{W}$ as:

\begin{equation}
    \mathbf{W} \approx \mathbf{U} \mathbf{V}^T
\label{eq:rank}
\end{equation}

where $\mathbf{U} \in \mathbb{R}^{m \times r}$ and $\mathbf{V} \in \mathbb{R}^{n \times r}$, with rank $r \ll \min(m, n)$. This leads to a reduction in the number of computations, FLOPs, and less power usage.














\textbf{Knowledge Distillation} Knowledge distillation is employed to transfer knowledge from a larger teacher model to a smaller student model to have a similar performance with fewer parameters. This process introduces a trade-off between the soft targets of the teacher and the ground truth labels, controlled by the parameter $\alpha$. The loss function in knowledge distillation is defined as:

\begin{equation}
L = \alpha L_{KD} + (1 - \alpha) L_{GT}
\end{equation}

where $L_{KD}$ is the distillation loss calculated using teacher soft predictions, and $L_{GT}$ is the loss using ground truth labels. A higher $\alpha$ puts more emphasis on the teacher’s outputs. In contrast, a lower $\alpha$ puts the student to rely more on the ground truth, which may limit the benefits of distillation. In summary, tuning $\alpha$ is to balance model efficiency.


\section{Proposed Approach}
To design an efficient image super-resolution model, we build upon a \new{Vision Mamba-based architecture} while optimizing its structure and using knowledge distillation. The goal is to maintain the performance of SR models while reducing computational complexity and model size and simplifying it for deployment on edge devices. By combining the efficient modeling of Mamba, embedding dimension adjustments, low-rank approximations, and tuning hyperparameters, MambaLiteSR achieves a balanced trade-off between model size, computational efficiency, and SR performance, making it suited for deployment on resource-constrained devices. In this section, we go over the main aspects and materials of our work. Experimental results show that this combination of parameters balancing allows for a substantial reduction in model size and energy usage while preserving high reconstruction accuracy.

Figure~\ref{high_level} presents our high-level diagram, which includes the model, knowledge distillation process, and parameters of interest. The flow consists of the following key components:

\begin{figure*}[ht]
	\centering
	\includegraphics[width=.9\textwidth]
         {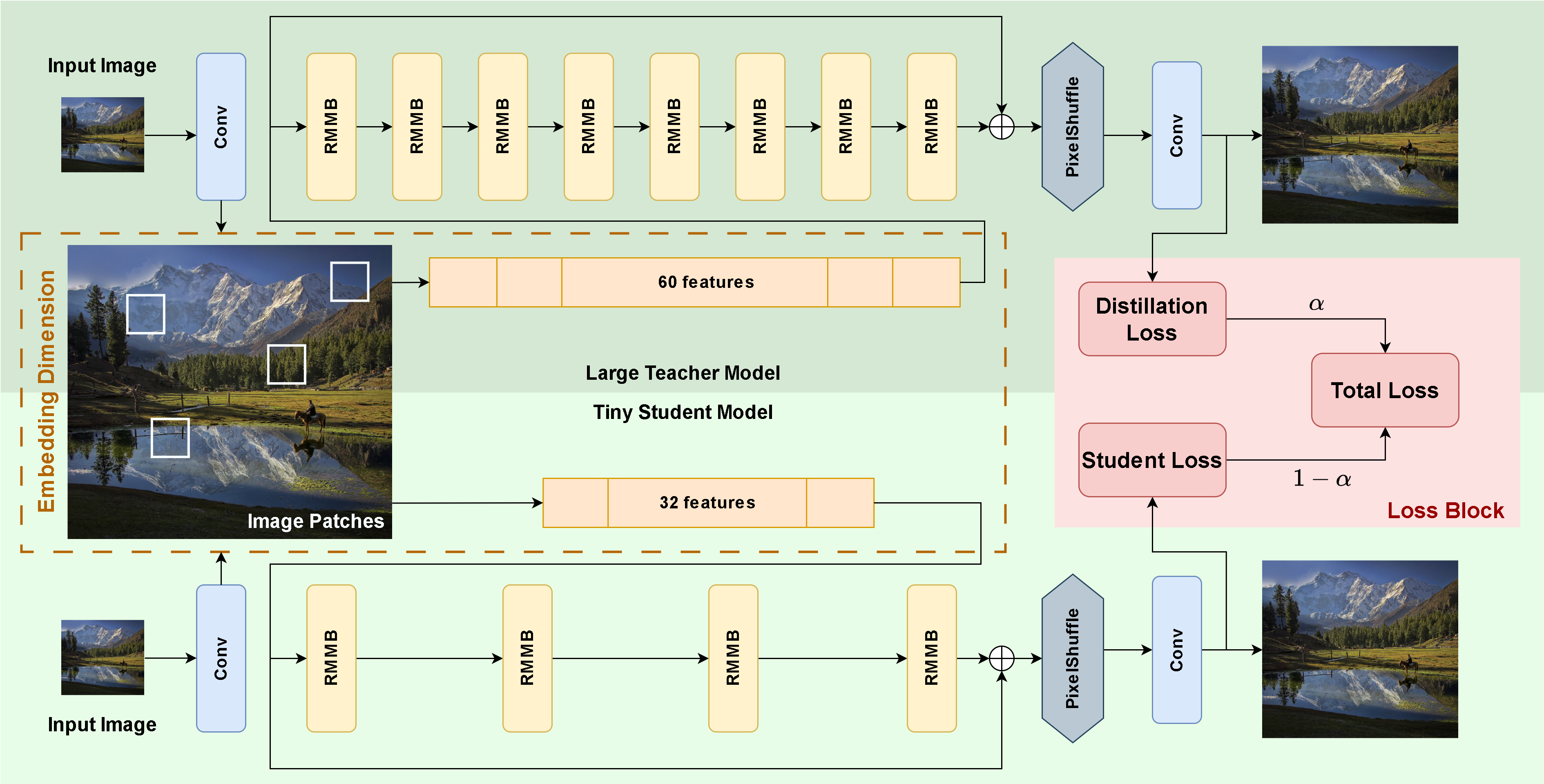}
	\caption{High-Level overview of MambaLiteSR process: Low-resolution input image~($64 \times 64$) is preprocessed and fed into the knowledge distillation process, generating the high-resolution output~($256 \times 256$). Weighted distillation and student losses enable the student model to learn efficiently under teacher supervision. Proper embedding dimension, which determines the feature vector size resulting from the image patches, makes it suitable for edge devices.}
	\label{high_level}
\end{figure*}

\subsection{Embedding Dimension}
The embedding dimension is a crucial factor in model efficiency, which directly affects the number of parameters and computational cost. By reducing the embedding dimension from 192 to 60 for the model, we achieve a substantial reduction in memory and power usage, which according to the Equation~\ref{eq:embed_dim}, leads to a 10 times reduction of the model size. Experimental results confirm that this reduction in embedding dimension does not significantly impact reconstruction quality and allows for efficient deployment on edge devices.

In this study, images are cropped into smaller patches. Specifically, from each image, eight mini-patches of $64 \times 64$ are extracted randomly. To further preprocess the data, random horizontal and vertical flips and rotations are applied. Then, each of these $64 \times 64$ patches is converted into a vector of 60 and 32 features, for teacher and student models respectively.

\subsection{Large Teacher Model}
The input image is first passed through a convolutional layer, which acts as a shallow feature extractor to derive basic feature maps. These features then go through a set of Residual Mixed Mamba Blocks~(RMMB), which consist of several Vision Mamba modules designed to capture long-range dependencies. Each RMMB internally contains a basic layer with a configurable number of Blocks, each Block has exactly one Mamba mixer inside and An internal residual convolution and skip connection as a residual group.

Therefore, RMMBs perform deep feature extraction and contain multiple Vision Mamba Modules along with a convolution layer and a residual connection. In the proposed large teacher model, the Mamba layer is low-rank to reduce the number of computations. After extracting deep features, a global residual connection is applied to fuse shallow features from the initial feature map with deep features. These features are then up-sampled by a method of pixel-shuffle to create the final HR output. To summarize, there are 16 Mamba layers in the teacher model with an embedding dimension of 60, which will be covered in the following subsections. 

\subsection{Tiny Student Model}
To improve efficiency, a student model is trained with a lower number of RMMBs and a smaller embedding dimension. The student model follows a similar architecture to the teacher but with reduced parameters. Knowledge distillation allows the student to mimic the performance of the teacher while being more computationally efficient. In this work, it has 4 RMMBs, resulting in 8 Mamba layers, with an embedding dimension of 32. The student model is optimized to keep essential features while achieving comparable PSNR and SSIM. Note that the input to both the teacher and student model is the same and they go through the same flow of preprocessing.

\subsection{Loss Block} As stated in the Equation~\ref{eq:alpha}, $\alpha$ controls the trade-off between distillation and direct supervision, and tuning it ensures the optimal balance between performance and model efficiency. In the total loss function, L1 loss for both distillation and student loss is considered. L1 loss measures the average difference between predicted and actual values. We write the combined loss as:

\begin{equation}
L = \alpha L_1(y_{student}, y_{teacher}) + (1 - \alpha) L_1(y_{student}, y_{GT})
\label{eq:alpha}
\end{equation}

By assigning a proper value for $\alpha$, each of the two loss terms is weighted to optimize the trade-off between performance and model size.

\subsection{Low-Rank Mamba}
Another key strategy is the low-rank approximation in the Mamba mixers. This reduces complexity and energy usage during training while maintaining model performance by factoring each $dim \times dim$ weight into $dim \times rank$ and $rank \times dim$. By reducing the rank from $dim / 2$, where $dim$ stands for the matrix dimension, we can ensure the reduction in computations. However, the rank layers are such a small fraction of the overall architecture, that changing ranks does not materially affect the total parameter count. But at runtime, the larger rank requires more FLOPs~(more multiplications/additions), so it draws more power under load. Therefore, rank mostly impacts computation rather than overall parameter storage. As a result, to reduce the energy usage and the number of parameters between layers, a low-rank approximation is applied.

In this work, we fine-tune these two hyperparameters: the distillation weight $\alpha$ and the rank used in low-rank approximation. These parameters are adjusted based on experimental evaluations to achieve an optimal trade-off between efficiency and performance.

\section{Experimental Result}
In this section, we begin by presenting the experimental setup, which includes details about the dataset and the implementation. Then, we evaluate the proposed LightMambaSR model, focusing on its size, performance, and power consumption in comparison to state-of-the-art methods.


\subsection{Experimental Setup}

\textbf{Datasets.} \new{In this work, we used the DF2K~(DIV2K + Flickr2K)~\cite{div2k} dataset, which consists of 3450 high-resolution images and their corresponding low-resolution versions. The low-resolution images are generated by downscaling each image by factors of 2, 3, and 4.}
\new{To train the proposed model, we consider a scale of 4 to do the experiments as it presents a more challenging problem and requires more parameters to learn. Moreover, for the validation set, a subset of the  DIV2K dataset consisting of 100 images was used. For testing the accuracy and performance of the proposed approach, we utilized four standard benchmark datasets: Set5~\cite{set5}, Set14~\cite{set14}, BSD100~\cite{bsd100}, and Urban100~\cite{urban100}.}

\textbf{Implementation Details.} For faster processing during training, images are cropped into smaller patches. Specifically, from each \new{LR image eight mini-patches of $64~\times~64$ and from each HR image eight corresponding mini-patches of $256~\times256$} are extracted randomly. To further preprocess the data, random horizontal and vertical flips and rotations are applied. The model is trained using the Adam optimizer. The batch size is set to 128 and the training process takes 2500 iterations. The initial learning rate is set at~$2 \times 10^{-4}$ and is halved when the training iteration reaches specific milestones. For the embedding dimension of images, it is considered to be 60 to reduce the model size. For evaluation, we calculate PSNR and SSIM metrics on the Y channel in the YCbCr color space. All \new{training} experiments are conducted on Lambda GPU Server~\cite{lambda}, which utilizes NVIDIA GeForce RTX 4090 GPU.

Experiments are conducted with various reduced matrix ranks including ranks 30 and 2. To further reduce the model size, knowledge distillation is applied, the same process as DVMSR~\cite{lei2024dvmsr}, but with different values and hyperparameters. For this purpose, the embedding dimension is 
reduced to 32. 
To assess the performance of different values of $\alpha$ parameter in knowledge distillation, an experiment was conducted comparing the result of changing this value over 1000 iterations for each of the teacher and student models.

\begin{figure}
	\centering
	\includegraphics[width=.45\textwidth]
         {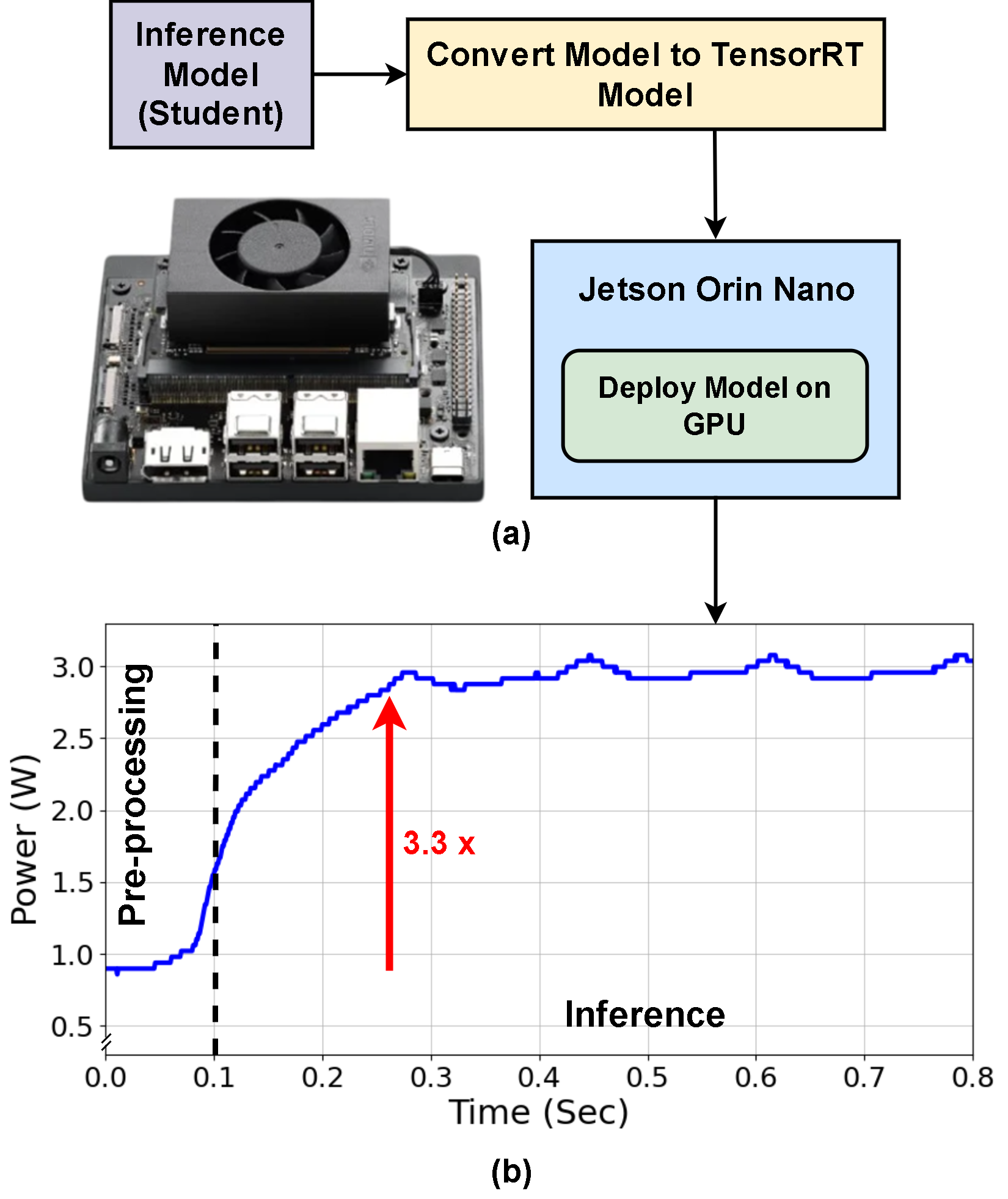}
	\caption{(a) For measuring dynamic power on the embedded NVIDIA Jetson Orin Nano, the student onnx model is converted to TensorRT format and then the measurement starts. (b) The plot shows the instantaneous power usage over time for 1000 samples, extracted using the tegrastats utility~\cite{tegrastats}. 
    }
	\label{instant_power}
\end{figure}

For the \new{inference} power consumption measurements, we deploy the model on the embedded Nvidia Jetson Orin Nano board~~\cite{orin} as shown in Figure~\ref{instant_power}.(a). \new{Power measurements experiments are repeated over 1000 samples, and the average is calculated. The Jetson Orin Nano Developer Kit with 8GB of memory features a hexa-core ARM Cortex-A78AE CPU with 1.5 GHz frequency and a 512-core NVIDIA Ampere GPU with 625 MHz. This board is optimized for AI tasks like matrix multiplications and deep learning inference, making it ideal for edge AI applications~\cite{vitreg}, besides boards such as Jetson Orin Nano~\cite{nano}.} We converted the software models to ONNX format and optimized them into TensorRT representations for GPU-accelerated inference. \new{The frequency was set to 612~MHz with the power mode being set to 7~Watt.} During inference, the tegrastats utility \new{\cite{tegrastats}} was used to measure instantaneous power. 

\subsection{Results} 

\new{In this section, we compare the proposed MambaLiteSR model with the state-of-the-art methods, including eSR~\cite{esr}, ESPCN~\cite{shi2016real}, and FSRCNN~\cite{dong2016accel}, in terms of performance and power consumption. While ESPCN and FSRCNN were not originally designed for edge deployment, they were evaluated on edge devices in the experiments conducted in eSR. Next, we present the model optimization and training results to analyze the impact of various variables of knowledge distillation and Mamba architecture on training and overall performance. Moreover, we compare the proposed MambaLiteSR with the baseline~\cite{lei2024dvmsr}.
Finally, we demonstrate the performance of the proposed tiny student model by comparing the model output with the ground truth image to demonstrate its ability to achieve similar quality.}

\begin{table*}[ht]
  \caption{Comparison of image quality and performance metrics of MambaLiteSR to existing methods implemented on the edge device Jetson AGX Xavier GPU~\cite{xavier}. 
  MambaLiteSR improves the power consumption by up to \new{58\%} to implement while performing similarly. The value of the power of the proposed approach is specific to the embedded NVIDIA Jetson Orin device. The reported power is the average of 1000 samples. Experiments were conducted on a scale of 4.}
  \label{tab:result}
  \centering\medskip
\resizebox{\linewidth}{!}{
  \begin{tabular}{lcclrrccccccccc}\hline
    \multirow{2}{*}{Algorithm} & Power & \multicolumn{2}{c}{Set5} & \multicolumn{2}{c}{Set14} & \multicolumn{2}{c}{BSDS100} & \multicolumn{2}{c}{Urban100} \\
                             & $\text{[mWatts]}$ & PSNR & SSIM & PSNR & SSIM & PSNR & SSIM & PSNR & SSIM \\ \hline
    eSR~\cite{navarrete2022edge} 
    & 7100 & 30.62 & 0.860 & 27.48 & 0.751 & 26.93 & 0.714 & 24.42 & 0.718\\
    eSR - fast~\cite{navarrete2022edge} 
    & 3867 & 28.64 & 0.806 & 26.12 & 0.712 & 26.13 & 0.684 & 23.28 & 0.668\\
    ESPCN~\cite{shi2016real} 
    & 6952 & 30.57 & 0.858 & 27.50 & 0.752 & 26.92 & 0.715 & 24.42 & 0.718\\
    FSRCNN~\cite{dong2016accel} 
    & 4795 & 30.16 & 0.845 & 27.19 & 0.742 & 26.74 & 0.707 & 24.09 & 0.702\\
    MambaLiteSR (Ours)
    & \new{\textbf{2957}} & 28.28 & 0.837 & 25.39 & 0.727 & 25.20 & 0.693 & 22.57 & 0.686 \\
    \noalign{\smallskip}\hline\noalign{\smallskip}
  \end{tabular}
}
\end{table*}

Figure~\ref{instant_power}.(b) illustrates the power consumption results on the embedded NVIDIA Jetson Orin Nano during inference. Experiments are done on the \new{proposed MambaLiteSR} tiny student model to evaluate the power and PSNR/SSIM results, as summarized in Table~\ref{tab:result}. The idle power consumption was measured at approximately \new{0.9~W}, and then, the system initiated image pre-processing. Following this, the model execution and image input processing caused a \new{3.3}$\times$ increase in power consumption. 
The deployment results of the proposed MambaLiteSR model, compared against state-of-the-art methods on edge devices, are detailed in Table~\ref{tab:result}. The power measurement was averaged over 1000 samples, \new{resulting in a fast single image latency of 14~ms. The results demonstrate that the proposed approach improved the power consumption by up to  
58\% 
while maintaining SSIM and PSNR performance comparable to existing methods. \new{Note that the compared methods are all CNN-based. While CNNs extract structured features, generative AI models such as ours can create more detailed and realistic images by generating missing information. Such models handle complex transformations better, especially for large scaling factors, and produce sharper textures by learning perceptual features instead of just minimizing pixel differences.}}



To reduce the number of parameters to make the model suitable for edge deployment the embedding dimension of \new{60 and 32 to the teacher and student model is applied respectively}, which leads to the 15$\%$ reduction in the number of parameters while keeping the performance acceptable with comparison with the Baseline~\cite{lei2024dvmsr}. The proposed LightMambaSR includes 370k parameters with a PSNR of 28.28 while the baseline has a 424k parameters with a PSNR of 32.19. There is a difference in PSNRs, which due to the increasing trend of the training we expect to achieve the same baseline PSNR as they did the experiments for 500K iterations as shown in Table~\ref{tab:dvmsr}.

\begin{table}[h]
    \centering
    \caption{Comparison of the number of parameters and PSNR usage for the baseline and MambaLiteSR at a scale factor of 4. The PSNR results are evaluated on the Set5 dataset.}  
    \begin{tabular}{llll}
        \hline
        Model & Parameters & iterations & PSNR  \\
        \hline
        DVMSR~\cite{lei2024dvmsr}  & 424k & 500,000 & 32.19\\
        MambaLiteSR  & \textbf{370k (15\% smaller)} & 2500 & 28.28\\
        \hline
    \end{tabular}
    \label{tab:dvmsr}
\end{table}


Table~\ref{tab:rank_batch} presents the rank configurations of MambaLiteSR along with their results. It can be seen that there is a negligible difference between the PSNRs of $rank = 2$ and $rank = 30$. As a result, $rank = 2$ is chosen to allow for 1.7x less energy consumption besides good performance.

\begin{table}[ht]
    \centering
    \caption{Comparison of PSNR and power usage for different ranks at a scale factor of 4. Configs 1 and 2 use the same configuration, varying only in rank and belonging to the teacher model. It can be seen that the lower-rank model consumes around 42\% less power than the higher-rank model. The reported power measurements are averaged values obtained during the training of the teacher model on the Lambda GPU Server, and the PSNR results are evaluated on the validation dataset.}  
    \begin{tabular}{llllll}
        \hline
        Config. & Rank & Parameters & iterations & PSNR & Power (W) \\
        \hline
        Config 1   & \textbf{30} & 370k & 1500 & 28.88 & \textbf{103}\\
        Config 2   & \textbf{2} & 370k & 1500 & 28.81 & \textbf{60} \textbf{(1.7x less)}  \\
        \hline
    \end{tabular}
    \label{tab:rank_batch}
\end{table}

\begin{figure*}[ht]
	\centering
	\includegraphics[width=.9\textwidth]
         {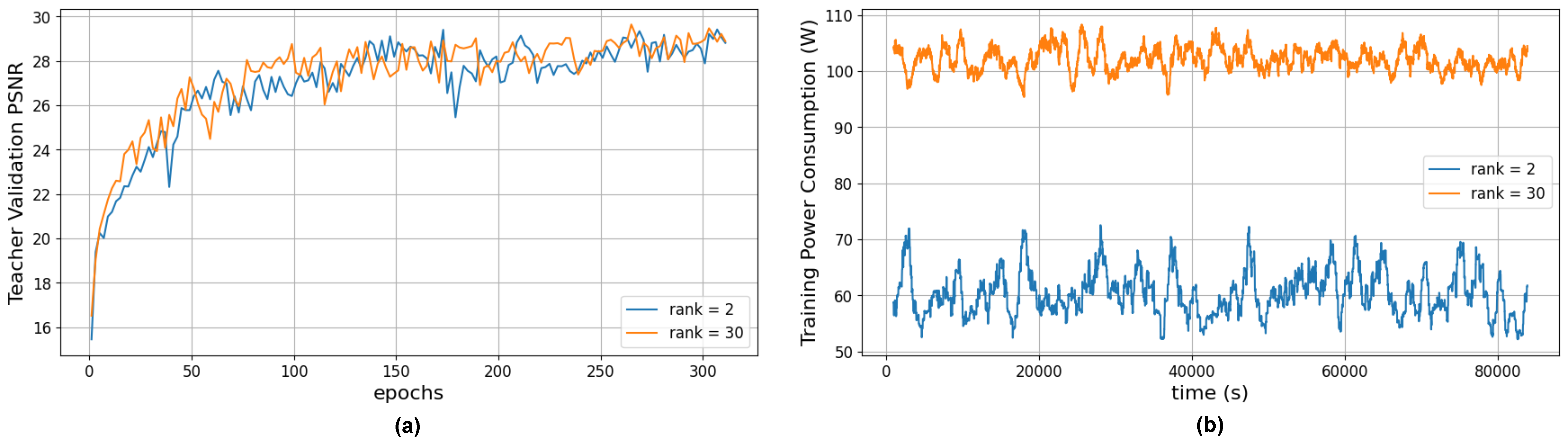}
	\caption{Comparison between MambaLiteSR teacher model training when $rank = 2$ and $rank = 30$ over 1500 iterations: (a) depicts the validation PSNR. (b) depicts the moving average of GPU power usage every 100 seconds with a window size of 100. At runtime, the larger rank requires more FLOPs, drawing more power under load. The training outcome and model performance act similar because, as indicated by Equation~\ref{eq:rank}, the matrix ultimately remains the same. The reported power measurements correspond to the NVIDIA GeForce RTX 4090 on Lambda GPU Server~\cite{lambda} using wandb~\cite{wandb} dashboard.}
	\label{ranks}
\end{figure*}

The performance of different values of rank can be seen in Figure~\ref{ranks}, which shows the validation results during teacher model training for rank values of 2 and 30, along with the moving average of their corresponding GPU power usage during training on the Lambda GPU Server~\cite{lambda}, with all other settings kept the same. While both models demonstrate similar training behavior and performance, the model with the lower rank demonstrates more efficient training; when $rank = 2$, GPU power usage is 42\% less than the one with $rank = 30$.

To further reduce the model size, knowledge distillation is applied. According to the Equation~\ref{eq:alpha}, $\alpha$ is a key parameter in the loss function. Therefore, experiments were made with different values of $\alpha$ to evaluate their performance over 1000 iterations and choose the most appropriate one. The result of different values of $\alpha$ can be seen in table~\ref{tab:alpha} with their corresponding PSNR.

\begin{table}[ht]
    \centering
    \caption{Comparison of PSNR for different $\alpha$ values. The reported PSNR belongs to the student model on the validation set over 1000 iterations.}  
    \begin{tabular}{c|ccccc}
        \hline
        \textbf{$\alpha$} & 0.2 & 0.4 & 0.6 & 0.8 \\
        \hline
        \textbf{PSNR} & 28.71 & 26.35 & 26.69 & 27.95 \\
        \hline
    \end{tabular}
    \label{tab:alpha}
\end{table}

\begin{figure}[t]
    \centering
    \includegraphics[width=0.9\linewidth]{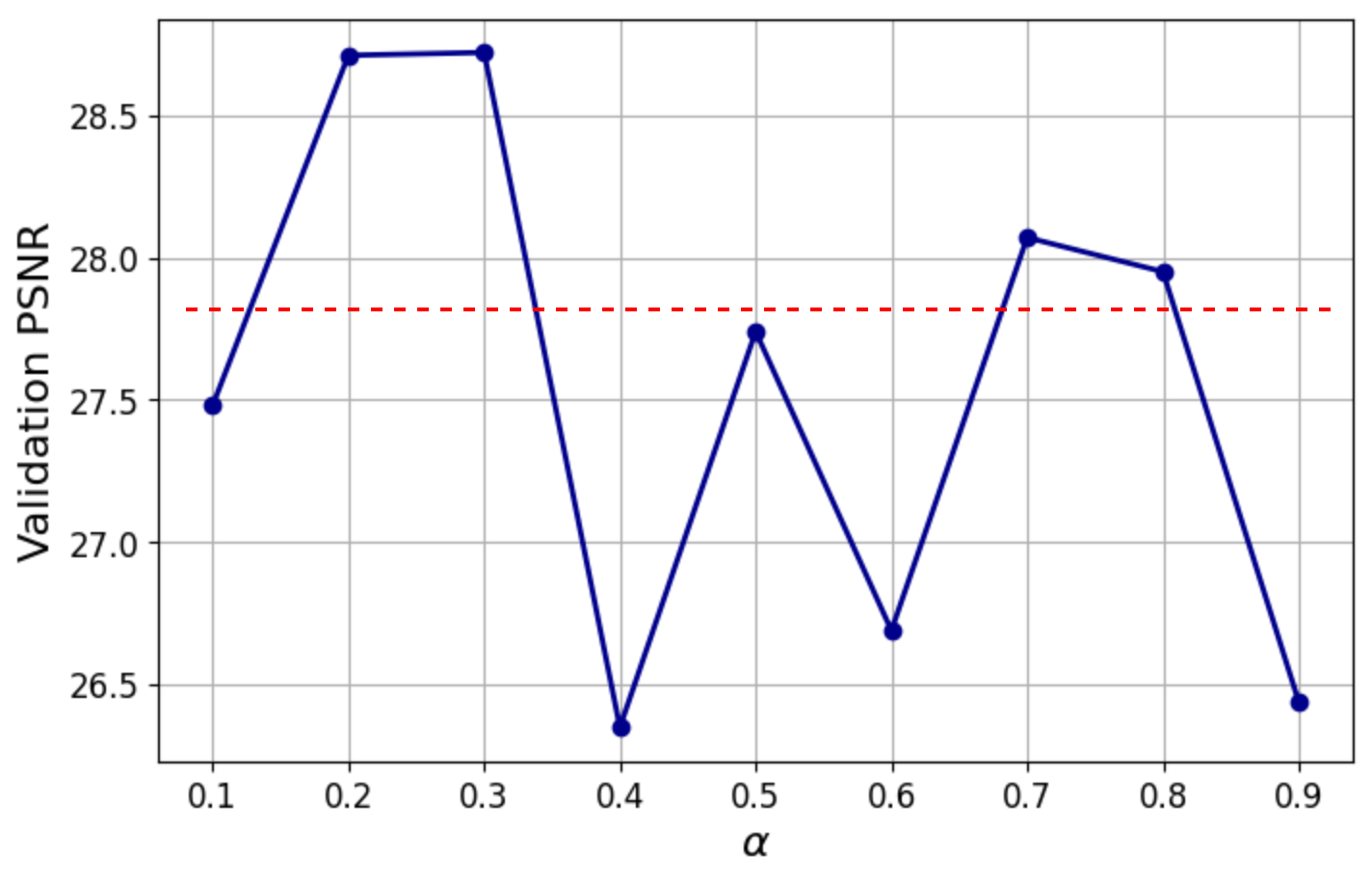}
    \caption{Model performance on the validations set based on the changes in $\alpha$, suggesting a potential inconsistency in gradient values between the learned teacher and the ground truth. The reported PSNRs are after 1000 iterations of training for teacher and student models.}
    \label{alpha}
\end{figure}

As a result, $\alpha = 0.8$ is chosen to enable both learning from the teacher and ground truth labels. After that, the student model is put to training for 2500 number of iterations, the same as the teacher model. Then, the experiments on $\alpha$ values and their corresponding PSNRs inspected more with a wider range of values because of the probable formulation it suggested. Figure~\ref{alpha} demonstrates model performance on the validation set with changes in $\alpha$, suggesting a potential inconsistency in gradient values between the learned teacher and the ground truth, as indicated in the Equation~\ref{eq:alpha}, meaning that whenever the teacher and ground truth have near to similar weights in the loss function, PSNR might degrade. Moreover, if one of these weights decreases significantly, it indicates that the model is either not effectively learning or has reverted to learning from scratch. \new{Note that in the baseline, $\alpha$ was set to 0.5, meaning the impact of the teacher model and ground truth is the same, which as was mentioned, results in inferior performance.}

\begin{figure}
	\centering
	\includegraphics[width=.45\textwidth]
         {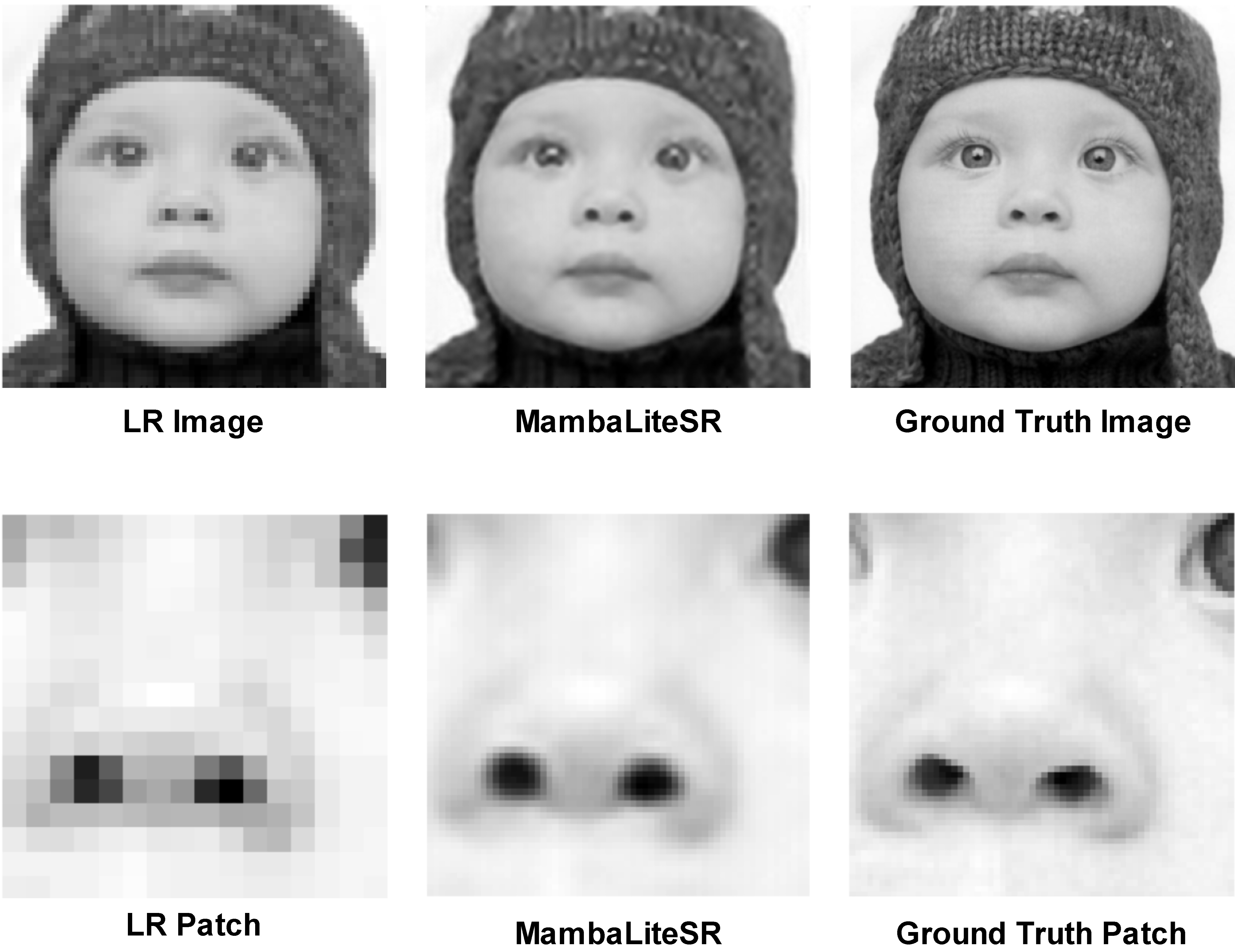}
	\caption{Demonstration of LR image, MambaLiteSR output, and ground truth image, as well as a random patch cropped from the LR input and its corresponding MambaLiteSR output and ground truth patch.}
	\label{patch}
\end{figure}

Figure~\ref{patch} demonstrates the preprocessed input image, the corresponding MambaLiteSR output, and the ground truth image. The same process for a random patch is shown as well for better visualization\new{; a random patch from the LR image is taken, was given as an input to the model, and the output as well as the ground truth is shown.} A complete process consists of dividing the LR image into patches, passing each of them through the model, and stitching these patches together. It can be seen that MambaLiteSR is successful in generating a super-resolution image similar to its ground truth image.

\section{Conclusion}
In this paper, we presented MambaLiteSR, a new image super-resolution model designed for edge devices. It combines a low-rank Mamba mixer with knowledge distillation to deliver outputs with super-resolution, relying on state space representations while checking the embedding dimension and rank settings to minimize parameters and computation, as well as the distillation parameter tuning for better performance. By integrating knowledge distillation, MambaLiteSR inherits the performance of a larger teacher model and achieves competitive PSNR and SSIM on standard benchmarks. Our experiments show that MambaLiteSR can reduce the model size by up to 15\% compared to a baseline without sacrificing reconstruction quality significantly. Changes in rank resulted in a 42\% reduction in power usage during training. Assessment of the distillation parameter led us to a realization between the teacher model and ground truth, resulting in a balanced knowledge distillation. We also demonstrated its real-time capability on the embedded NVIDIA Jetson Orin Nano, which maintains \new{lower power consumption compared to the state-of-the-art edge super-resolution models, while maintaining similar performance}. By combining low-rank approximation, knowledge distillation, and efficient embedding strategies, MambaLiteSR offers a practical solution for generative AI applications on edge devices.


\section{Acknowledgment}
This project was sponsored by the U.S. Army Research Laboratory.

\bibliographystyle{IEEEtran} 
\bibliography{references}    

\end{document}